\documentclass{easychair}

\usepackage{doc}
\usepackage{tikz}
\usetikzlibrary{arrows.meta}
\usetikzlibrary{shapes.multipart}
\usepackage{amsfonts}
\usepackage[linesnumbered]{algorithm2e}

\title{A typo in the Paterson-Wegman-de~Champeaux algorithm}

\author{
Valeriu Motroi\inst{1}
\and
{\c S}tefan Ciob{\^ a}c{\u a}\inst{2}
}

\institute{
  Alexandru Ioan Cuza University
  Ia{\c s}i, Romania\\
  \email{\{motroival, stefan.ciobaca\}@gmail.com}
}

\authorrunning{Motroi and Ciob{\^ a}c{\u a}}

\titlerunning{A typo in the Paterson-Wegman-de Champeaux algorithm}

\begin{document}

\maketitle

\begin{abstract}
  We investigate the Paterson-Wegman-de~Champeaux linear-time
  unification algorithm. We show that there is a small mistake in the
  de~Champeaux presentation of the algorithm and we provide a fix.
\end{abstract}
\section{Introduction}

In this paper we investigate the Paterson-Wegman
algorithm~\cite{paterson1976}, as improved by de
Champeaux~\cite{champeaux1986}. The algorithm has linear-time
complexity. In Figure~\ref{mainalg} we present the pseudo-code
proposed by de~Champeaux. We add line numbers and we make some
cosmetic changes, which do not affect the algorithm logic. For
example, we omit the \emph{else} branch of an \emph{if} statement
whose \emph{then} branch ends with an \emph{exit} statement. The
de~Champeaux presentation of the algorithm ends with a post-processing
step, described in Figure~\ref{postprocess}.

The issue we identify is that the post-processing step enters an
infinite loop. The infinite loop is caused by a bug in the
occurs-check test. We give an input producing an infinite loop in the
next section. The bug can be fixed syntactically by indenting an
assignment statement, i.e., moving it inside the inner code block.

This issue was noticed and fixed by Erik~
Jacobsen~\cite{jacobsen1991unification} (see footnote on Page
34). However, in this paper we present and analyze an troublesome
input in detail.

\begin{figure} 
\begin{algorithm}[H]
\SetKwFunction{Finish}{Finish}
\SetKwFunction{Solver}{Solver}
\SetKwFunction{BUILDSIGMA}{BUILD-SIGMA}
\SetKwFunction{EXPLOREVARIABLE}{EXPLORE-VARIABLE}
\SetKwFunction{DESCEND}{DESCEND}
\SetKwFunction{EXPLOREARGUMENTS}{EXPLORE-ARGUMENTS}
\SetKwProg{Fn}{Function}{:}{}
\SetKwProg{Prcd}{Procedure}{:}{}
\SetKwFor{For}{for (}{)}{}
\SetKwFor{Foreach}{FOR-EACH }{do}{}
\Prcd{\Solver{u, v}} {
  Create link (u, v)\\
  \textbf{While} there is a function node r, \textbf{Finish}(r)\\
  \textbf{While} there is a variable node r, \textbf{Finish}(r)\\
  \textbf{BUILD-SIGMA}(SIGMA)
}
\Prcd{\Finish{r}} {
  \uIf{complete(r)} {
    Exit
  }
  \uIf{pointer(r) $\neq$ NIL} {
    Exit with failure
  }
  Create new pushdown stack with operations \textbf{Push}(*) and \textbf{Pop}\\
  pointer(r) := r\\
  \textbf{Push}(r)\\
  \While{stack $\neq$ NIL} {
    s := \textbf{Pop}\\
    \uIf{r, s have different function symbols} {
      Exit with failure
    }
    \Foreach{parent t of s} {
      \textbf{Finish}(t)
    }
    \Foreach{link (s, t)} {
      \uIf{Complete(t) or t = r} {
        \textbf{Ignore} t
      }
      \uElseIf{pointer(t) = NIL} {
        pointer(t) := r\\
        \textbf{Push}(t)
      }
      \uElseIf{pointer(t) $\neq$ r} {
        Exit with failure
      }
      \uElse {
        Ignore t // (since t is already on STACK)
      }
    }
    \uIf{s $\neq$ r} {
      \uIf{Variable(s)} {
        Subs(s) := r\\
        Add s to SIGMA (input to BUILD-SIGMA)
      }
      \uElse {
        Create links\{\textit{j}th son(r), \textit{j}th son(s) $|$ $1 \leq j \leq q$\}
      }
    }
    Complete(s) := true
  }
  Complete(r) := true
}
\end{algorithm}
\caption{
    \label{mainalg} Paterson-Wegman algorithm as presented by de~Champeaux. We add line numbers and we make some cosmetic changes.
}
\end{figure}

\section{Troublesome example}

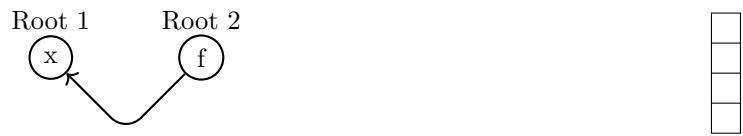
\begin{figure}
\begin{minipage}[b]{0.45\textwidth}
\begin{center}
\begin{tikzpicture}
\begin{scope}[every node/.style={circle,thick,draw}]
  \node (f) at (1, 0) {f};
  \node (x) at (-1, 0) {x};
\end{scope}

\begin{scope}
 \node (root1) at (-1, 0.5) {Root 1};
 \node (root2) at (1, 0.5) {Root 2};
\end{scope}

\begin{scope}[thick,rounded corners=8pt, ->]
  \draw (f) -- (0, -1) -- (x);
\end{scope}
\end{tikzpicture}
\end{center}
\end{minipage}
\hfill
\begin{minipage}[b]{0.45\textwidth}
\begin{center}
 \begin{tikzpicture}[stack/.style={rectangle split, rectangle split parts=#1,draw, anchor=center}]
\node[stack=4]  {
};
\end{tikzpicture}
\end{center}
\end{minipage}

\caption{
    \label{state1} The data structures at the start of the algorithm.
}
\end{figure}

We show how the de~Champeaux algorithm works when trying to unify the
terms $X$ and $f(X)$. The algorithm starts with the DAG representation
of the two terms, which we show in Figure~\ref{state1}. As the two
terms have maximal sharing between them, there is only one node
labeled $X$. There are two roots, each corresponding to one of the
terms to be unified. We use simple arrows to denote the relation
between parent and child nodes of the DAG.

The algorithm creates \emph{links} (undirected edges) between nodes
that should be in the same equivalence relation. We use dashed lines
to denote the links created by the algorithm. The algorithm also
maintains stacks (shown graphically on the right) and a set of
pointers from nodes to nodes, which are represented by two-headed
arrows.

The algorithm starts by creating a link between $X$ and $f(X)$ (Figure~\ref{state2}).
\begin{figure}
\begin{minipage}[b]{0.45\textwidth}
\begin{center}
\begin{tikzpicture}
\begin{scope}[every node/.style={circle,thick,draw}]
  \node (f) at (1, 0) {f};
  \node (x) at (-1, 0) {x};
\end{scope}

\begin{scope}[thick,rounded corners=8pt, ->]
  \draw (f) -- (0, -1) -- (x);
\end{scope}

\begin{scope}[thick, rounded corners=8pt, -, dashed]
  \draw (x) -- (f);
\end{scope}

\end{tikzpicture}
\end{center}
\end{minipage}
\hfill
\begin{minipage}[b]{0.45\textwidth}
\begin{center}
 \begin{tikzpicture}[stack/.style={rectangle split, rectangle split parts=#1,draw, anchor=center}]
\node[stack=4]  {
};
\end{tikzpicture}
\end{center}
\end{minipage}
\caption{\label{state2} The data structures representation after adding the first link.}
\end{figure}

The next step is to call \emph{Finish} on all functional nodes (line 3). In this example we have only one functional node, $f$.
At this step, we have $r = f(X)$. Because \emph{complete(r)} is marked as \emph{false} and \emph{pointer(r)} is \emph{NIL}, 
we jump straight to line 12, where we set \emph{pointer(r)} to \emph{r} and push it to the stack (Figure~\ref{state3}).

\begin{figure}
\begin{minipage}[b]{0.45\textwidth}
\begin{center}
\begin{tikzpicture}
\begin{scope}[every node/.style={circle,thick,draw}]
  \node (f) at (1, 0) {f};
  \node (x) at (-1, 0) {x};
\end{scope}

\begin{scope}[thick,rounded corners=8pt, ->]
  \draw (f) -- (0, -1) -- (x);
\end{scope}

\begin{scope}[thick, rounded corners=8pt, -, dashed]
  \draw (x) -- (f);
\end{scope}

\begin{scope}[thick, rounded corners=8pt, ->>]
  \draw (f) -- (2, 0.7) -- (2, 0.2) -- (f);
\end{scope}

\end{tikzpicture}
\end{center}
\end{minipage}
\hfill
\begin{minipage}[b]{0.45\textwidth}
\begin{center}
 \begin{tikzpicture}[stack/.style={rectangle split, rectangle split parts=#1,draw, anchor=center}]
\node[stack=4]  {
\nodepart{four}f(X)
};
\end{tikzpicture}
\end{center}
\end{minipage}
\caption{\label{state3} The data structures after pushing the first functional node to the stack.}
\end{figure}

At the first iteration of the while loop, at line 15, we have $s = r$. As $s$ and $r$ have the same function symbol, we do not enter the if statement at line 16. As \emph{s} does not have any parent, we do not enter the if statement at line 18. 
The variable $s$ has a link to $X$ and, as a result, at line 21 we have $r = f(X)$, $s = f(X)$, $t = X$. The variable $t$ is not marked complete and is not equal to 
$r$, so we enter the if statament at line 24, set \emph{pointer(t)} to be $r$ and push it on the stack (Figure~\ref{state4}).
\begin{figure}
\begin{minipage}[b]{0.45\textwidth}
\begin{center}
\begin{tikzpicture}
\begin{scope}[every node/.style={circle,thick,draw}]
  \node (f) at (1, 0) {f};
  \node (x) at (-1, 0) {x};
\end{scope}

\begin{scope}[thick,rounded corners=8pt, ->]
  \draw (f) -- (0, -1) -- (x);
\end{scope}

\begin{scope}[thick, rounded corners=8pt, -, dashed]
  \draw (x) -- (f);
\end{scope}

\begin{scope}[thick, rounded corners=8pt, ->>]
  \draw (f) -- (2, 0.7) -- (2, 0.2) -- (f);
  \draw (x) -- (-2, 0.7) -- (-2, 0.2) -- (x);
\end{scope}

\end{tikzpicture}
\end{center}
\end{minipage}
\hfill
\begin{minipage}[b]{0.45\textwidth}
\begin{center}
 \begin{tikzpicture}[stack/.style={rectangle split, rectangle split parts=#1,draw, anchor=center}]
\node[stack=4] {
\nodepart{four}X
};
\end{tikzpicture}
\end{center}
\end{minipage}
\caption{\label{state4} The data structures after adding the variable X to the stack.}
\end{figure}

After this step, we jump straight to line 30, because there is only one link. 
We do not enter the if statement at line 30 because $s$ equals $r$. Then we set \emph{complete(s)} to \emph{true} at line 36.
Note that $s$ is still $f$. In the next iteration of the while loop at line 14 we have $s = X$. 
Because of the shared structure of common variables, we call \emph{Finish($f$)} at line 19, but \emph{complete($f$)} is true, 
so we exit this function call at line 8. Next follows the loop at line 20. We have the initial link $X$ and $f(X)$, so in this 
case $t = f(X)$, but \emph{complete(t)} is true and the node $t$ is ignored (line 22). Moving on, on line 30, we enter the if statement and jump to line 32, 
because $s = X$, which is a variable. At line 32 we set $subs(X) = f(X)$ and at line 33 we add $X$ to \emph{SIGMA}. Then, at line 36, we set \emph{complete(s)} to true.
The stack is now empty, so we go to the line 38 where we set \emph{complete(r)} to true (this is the second time we set \emph{complete(s)} to true). The execution of \emph{Finish} is done and we call \emph{Finish} on all variable nodes. We have only one variable, $X$, which has \emph{complete(X)} set to true, so we immediately return. Now we call \emph{BUILD-SIGMA}. One important observation is that we finished the main algorithm and the occurs-check at line 9 did not happen.

In Figure~\ref{postprocess} we show the implementation of \emph{BUILD-SIGMA}.
\begin{figure}
 \begin{algorithm}[H]
\SetKwFunction{Finish}{Finish}
\SetKwFunction{Solver}{Solver}
\SetKwFunction{BUILDSIGMA}{BUILD-SIGMA}
\SetKwFunction{EXPLOREVARIABLE}{EXPLORE-VARIABLE}
\SetKwFunction{DESCEND}{DESCEND}
\SetKwFunction{EXPLOREARGUMENTS}{EXPLORE-ARGUMENTS}
\SetKwProg{Fn}{Function}{:}{}
\SetKwProg{Prcd}{Procedure}{:}{}
\SetKwFor{For}{for (}{)}{}
\SetKwFor{Foreach}{FOR-EACH }{do}{}

\Prcd{\BUILDSIGMA{list-of-variables}} {
  \Foreach{variable $x_i$ in list-of-variables} {
    Add to final substitution $x_i \rightarrow $ \textbf{EXPLORE-VARIABLE($x_i$)} 
  }
}

\Fn{\EXPLOREVARIABLE{$x_i$}} {
  \uIf{Ready($x_i$) $\neq$ NIL} {
    Exit with Ready($x_i$)
  }
  out := \textbf{DESCEND(Subs($x_i$))} \\
  \uIf{out = NIL} {
    out := $x_i$
  }
  Ready($x_i$) := out\\
  Exit with out
}

\Fn{\DESCEND{$u_i$}} {
  \uIf{$u_i$ = NIL} {
    Exit with NIL
  }
  \uIf{Variable($u_i$)} {
    Exit with \textbf{EXPLORE-VARIABLE($u_i$)}
  }
  \uIf{Constant($u_i$)} {
    Exit with $u_i$
  }
  \uIf{Ready($u_i$)} {
    Exit with Ready($u_i$)
  }
  out := \textbf{EXPLORE-ARGUMENTS}(arguments-of($u_i$)) \\
  \uIf{out = arguments-of($u_i$)} {
    Ready($u_i$) := out
  }
  \uElse {
    // Cons gets as first argument a node and as a second argument \\ 
    // a pointer to a list of nodes and will return a pointer to\\
    // a list of nodes with the first argument in front \\
    // of the second argument. \\
    Ready($u_i$) := Cons(Head-of($u_i$), out)
  }
  Exit with Ready($u_i$)
}

\Fn{\EXPLOREARGUMENTS{list-of-arguments}} {
  \uIf{list-of-arguments = NIL} {
    Exit with NIL
  }
  1st-new := \textbf{DESCEND}(1st(list-of-arguments)) \\
  tail-new := \textbf{EXPLORE-ARGUMENTS}(tail(list-of-arguments))\\
  \uIf{1st-new $\neq$ 1st(list-of-arguments) OR \\tail-new $\neq$ tail(list-of-arguments)} {
    Exit with Cons(1st-new, tail-new)
  }
  Exit with list-of-arguments
}
\end{algorithm}
\caption{
    \label{postprocess} Post-processing step described by de~Champeaux.
}
\end{figure}
The function \emph{BUILD-SIGMA} creates a substitution from a ordered substitution in linear time. By running the algorithm, we conclude that it enters an infinite loop. In short, below are order of the function calls.

\begin{enumerate}
\item \emph{BUILD-SIGMA(list($X$))} - at line 1
\item \emph{EXPLORE-VARIABLE($X$)} -at line 3
\item \emph{DESCEND($f(X)$)} - at line 7
\item \emph{EXPLORE-ARGUMENTS(list($X$))} - at line 21
\item \emph{DESCEND($X$)} - at line 34
\item \emph{EXPLORE-VARIABLE}($X$) - at line 16
\end{enumerate}

The \emph{Ready} variable is not used. As a result, we enter a infinite loop.

\section{Fixing the de~Champeaux algorithm}

The issue with the pseudo-code presented by de~Champeaux is on line 36
in the \emph{Finish} procedure.  Based on the pseudo-code by
Paterson-Wegman, \emph{Complete(s)} should be set to true inside the
if statement at line 36. We propose a fixed version in
Figure~\ref{wrongvscorrectcode}.  This change fixes the pseudo-code
and the algorithm remains linear time and there are no further issues.

\begin{figure}[!tbp]
  \begin{minipage}[b]{0.49\textwidth}
    \begin{algorithm}[H]
      \setcounter{AlgoLine}{5}
      \SetKwFunction{Finish}{Finish}
      \SetKwFunction{Solver}{Solver}
      \SetKwFunction{BUILDSIGMA}{BUILD-SIGMA}
      \SetKwFunction{EXPLOREVARIABLE}{EXPLORE-VARIABLE}
      \SetKwFunction{DESCEND}{DESCEND}
      \SetKwFunction{EXPLOREARGUMENTS}{EXPLORE-ARGUMENTS}
      \SetKwProg{Fn}{Function}{:}{}
        \SetKwProg{Prcd}{Procedure}{:}{}
        \SetKwFor{For}{for (}{)}{}
        \SetKwFor{Foreach}{FOR-EACH }{do}{}
        \Prcd{\Finish{r}} {
        \uIf{complete(r)} {
            Exit
        }
        \uIf{pointer(r) $\neq$ NIL} {
            Exit with failure
        }
        Create new pushdown stack with operations \textbf{Push}(*) and \textbf{Pop}\\
        pointer(r) := r\\
        \textbf{Push}(r)\\
        \While{stack $\neq$ NIL} {
            s := \textbf{Pop}\\
            \uIf{r, s have different function symbols} {
            Exit with failure
            }
            \Foreach{parent t of s} {
            \textbf{Finish}(t)
            }
            \Foreach{link (s, t)} {
            \uIf{Complete(t) or t = r} {
                \textbf{Ignore} t
            }
            \uElseIf{pointer(t) = NIL} {
                pointer(t) := r\\
                \textbf{Push}(t)
            }
            \uElseIf{pointer(t) $\neq$ r} {
                Exit with failure
            }
            \uElse {
                Ignore t
            }
            }
            \uIf{s $\neq$ r} {
            \uIf{Variable(s)} {
                Subs(s) := r\\
                Add s to SIGMA
            }
            \uElse {
                Create links\{\textit{j}th son(r), \textit{j}th son(s) $|$ $1 \leq j \leq q$\}
            }
            }
            Complete(s) := true
        }
        Complete(r) := true
        }
    \end{algorithm}
  \end{minipage}
  \hfill
  \begin{minipage}[b]{0.48\textwidth}
    \begin{algorithm}[H]
      \setcounter{AlgoLine}{5}
    \SetKwFunction{Finish}{Finish}
    \SetKwFunction{Solver}{Solver}
    \SetKwFunction{BUILDSIGMA}{BUILD-SIGMA}
    \SetKwFunction{EXPLOREVARIABLE}{EXPLORE-VARIABLE}
    \SetKwFunction{DESCEND}{DESCEND}
    \SetKwFunction{EXPLOREARGUMENTS}{EXPLORE-ARGUMENTS}
    \SetKwProg{Fn}{Function}{:}{}
    \SetKwProg{Prcd}{Procedure}{:}{}
    \SetKwFor{For}{for (}{)}{}
    \SetKwFor{Foreach}{FOR-EACH }{do}{}
    \Prcd{\Finish{r}} {
    \uIf{complete(r)} {
        Exit
    }
    \uIf{pointer(r) $\neq$ NIL} {
        Exit with failure
    }
    Create new pushdown stack with operations \textbf{Push}(*) and \textbf{Pop}\\
    pointer(r) := r\\
    \textbf{Push}(r)\\
    \While{stack $\neq$ NIL} {
        s := \textbf{Pop}\\
        \uIf{r, s have different function symbols} {
        Exit with failure
        }
        \Foreach{parent t of s} {
        \textbf{Finish}(t)
        }
        \Foreach{link (s, t)} {
        \uIf{Complete(t) or t = r} {
            \textbf{Ignore} t
        }
        \uElseIf{pointer(t) = NIL} {
            pointer(t) := r\\
            \textbf{Push}(t)
        }
        \uElseIf{pointer(t) $\neq$ r} {
            Exit with failure
        }
        \uElse {
            Ignore t
        }
        }
        \uIf{s $\neq$ r} {
        \uIf{Variable(s)} {
            Subs(s) := r\\
            Add s to SIGMA
        }
        \uElse {
            Create links\{\textit{j}th son(r), \textit{j}th son(s) $|$ $1 \leq j \leq q$\}
        }
        Complete(s) := true
        }
    }
    Complete(r) := true
    }
    \end{algorithm}
  \end{minipage}
  \caption{\label{wrongvscorrectcode} On the left-hand side we present
    the initial pseudo-code due to de~Champeaux.On the right-hand side
    we propose the corrected version. The only difference is at line
    36 (note the indentation level).}
\end{figure}

\section{Conclusion}

We investigate the Paterson-Wegman linear-time unification algorithm
as improved by de~Champeaux. We show an example where the occurs-check
test fails to work as expected and results in an infinite loop in the
post-processing step. We show that the issue is caused by a
misindented statement (line 36) in the pseudo-code. Once the statement
is properly indented, the algorithm is correct and works in
linear-time as claimed.

\bibliography{refs}
\bibliographystyle{plain}

\end{document}